# Designing a Broadband Pump for High-Quality Micro-Lasers via Modified Net Radiation Method

Sergey Nechayev[1], Philip D. Reusswig[2], Marc. A. Baldo[2], Carmel Rotschild[1*]

[1]Department of Mechanical Engineering and Russell Berrie Nanotechnology Institute, Technion-Israel Institute of Technology, Haifa 32000, Israel

[2]Department of Electrical Engineering and Computer Science, Massachusetts Institute of Technology, 77 Massachusetts Avenue, Cambridge, MA 02139, USA

*Corresponding authors: carmelr@technion.ac.il

**High-quality micro-lasers are key ingredients in non-linear optics, communication, sensing and low-threshold solar-pumped lasers. However, such micro-lasers exhibit negligible absorption of free-space broadband pump light. Recently, this limitation was lifted by cascade energy transfer, in which the absorption and quality factor are modulated with wavelength, enabling non-resonant pumping of high-quality micro-lasers and solar-pumped laser to operate at record low solar concentration. Here, we present a generic theoretical framework for modeling the absorption, emission and energy transfer of incoherent radiation between cascade sensitizer and laser gain media. Our model is based on linear equations of the modified net radiation method and is therefore robust, fast converging and has low complexity. We apply this formalism to compute the optimal parameters of low-threshold solar-pumped lasers. It is revealed that the interplay between the absorption and self-absorption of such lasers defines the optimal pump absorption below the maximal value, which is in contrast to conventional lasers for which full pump absorption is desired. Numerical results are compared to experimental data on a sensitized $Nd^{3+}$:YAG cavity, and quantitative agreement with theoretical models is found. Our work modularizes the gain and sensitizing components and paves the way for the optimal design of broadband-pumped high-quality micro-lasers and efficient solar-pumped lasers.**

## INTRODUCTION

On-chip applications for sensing[1,2], non-linear optics[3,4] and optical communication[5,6] require high-quality factor (high-Q), micro-lasers. Also solar-pumped lasers[7] (SPLs) have similar demands due to the low solar flux density. Owing to the ultra-high transparency of a gain media and short cavity length of such lasers, the pump must propagate for many cavity cycles before being absorbed. Therefore, the coupling of non-resonant light to high-Q micro-lasers is inefficient. For SPLs[8–10], in addition to the mode coupling losses, the poor spectral overlap between the sun and the laser gain medium leads to high solar concentration at threshold, low slope efficiency and the need for solar tracking and active cooling[8–14]. Cascade energy transfer (CET) is a concept in which the absorption and emission spectra of materials form an energetic cascade[15–21]. CET pump schema enables broadband pumping of high-Q cavities[22] and SPLs that operate at record low solar concentrations[23]. Optimization of the radiative transfer in CET pump schema is rather challenging endeavor, requiring complicated numerical methods. Coherent methods are inapplicable due to the incoherent nature of excitation. Alternatively, Monte Carlo stochastic approach[24] is utilized to analyze an incoherent photon transport. However, if the optical path is large in the considered configuration, a calculation for even a single point of a parameter space is time-consuming. Moreover, such methods don't provide a physical intuition on the involved parameters. In this paper, we develop a theoretical model based on modified net radiation method[25] that includes a pump and CET sensitizer for planar waveguide[26,27] micro-lasers. This simplistic approach may also encompass CET sensitized micro-lasers in different geometries. Our model is implemented for solar-pumped lasers and for the specific configuration of $Nd^{3+}$:YAG cavity under ideal sensitization. Finally, we present our experimental observations of energy transfer from an organic sensitizer $AlQ_3$:DCJTB(2%):Pt(TPBP)(4%)[28] to an $Nd^{3+}$:YAG cavity, which shows excellent conformity with the net radiation model's prediction. It is revealed that such a sensitizer enables SPL to operate under non-concentrated sunlight, but the slope efficiency is limited to 0.53% due to optical losses. We discuss the advantages of such a generic and modular method for developing broadband-pumped high-quality micro-lasers.

## RESULTS

A schematic of the CET sensitized micro-laser in the slab configuration is shown in Figure **1a**. Incident light is absorbed by a layer of sensitizer and is then re-emitted as luminescence. As in the operation of a luminescent solar concentrator (LSC)[28–31], a fraction of the emitted photoluminescence is trapped within the waveguide formed by the optical gain media and its sensitizer coating[30]. Light propagating in the waveguide structure is subject to two competitive processes: the absorption in the gain media and the self-absorption in the sensitizer layer itself. Low power threshold $P_{th}$ micro-lasers must have a small mode volume $V$ because $P_{th}$ is proportional to the mode volume $V$ divided by the $Q$-factor: $P_{th} \propto V/Q$, the $Q$-factor being the ratio between the stored energy in the cavity to the energy dissipated per oscillation cycle. Lowering the mode volume means reducing the cavity thickness for the planar waveguide, but this change lowers the cavity absorption at the sensitizer emission wavelength, which must overcome the sensitizer self-absorption for adequate pumping.

Consider a sensitizing material with absorption constant $\alpha_a$ at pump wavelength. Efficient pump absorption requires that the sensitizer layer thickness must be $t_s \sim \alpha_a^{-1}$, which defines the self-absorption at the sensitizer emission wavelength to be $t_s \alpha_s \sim \alpha_s / \alpha_a$, where $\alpha_s$ is the sensitizer self-absorption constant at its emission wavelength. For effective pumping of the gain medium the sensitizer emission wavelength must match the gain media peak absorption coefficient $\alpha_g$. Moreover, cavity absorption must overcome the sensitizer self-absorption, i.e., $t_s \alpha_s \sim \alpha_s / \alpha_a \ll t_g \alpha_g$, where $t_g$ is the thicknesses of the gain medium. Consequently, while low power threshold $P_{th}$ demands small $t_g$, effective CET requires $t_g \gg \alpha_s / (\alpha_g \alpha_a)$. Optimization of this tradeoff between $P_{th}$ and the CET pumping efficiency is the key to designing an effective CET sensitizer for broadband-pumped high-quality micro-lasers.

We utilize the modified net radiation method[25], which is a convenient tool for addressing incoherent light absorption (see the Supplementary section I for details on absorption) in planar stratified media, to calculate the profile of the pump light absorption in the sensitizer and the sensitizer's luminescence absorption in the gain media. In our model, the solar-pumped micro-laser consists of *N* parallel layers (Figure **1b**), i.e., *N+2* geometric regions with semi-infinite free-space above the upper interface and below the lower interface. The indices *i=0* and *i=N* correspond to the upper and lower interfaces, respectively. Without loss of generality, the sensitizer is assumed to be the 1st layer of the micro-laser, and the gain media is the 3rd. Consider the media enclosed between planes *i* and *i+1,* as depicted in Figure **1b**. At the *i*th plane, the outgoing and incoming intensities are designated as $J_i^\pm(\omega,\theta)$ and $G_i^\pm(\omega,\theta)$, respectively, where $\omega, \theta$ stand for the angular frequency and angle of incidence, respectively. Planar systems have axial symmetry, and therefore, $\theta$ measured in any layer defines the angle in all of the other layers by Snell's law. The sign " $+$ " defines intensity components that are situated in the medium above the interface and the sign " $-$ " defines them when situated in the medium below the interface. In addition, in each medium, the incoming and outgoing intensity components are connected by the equations $G_i^+ = T_{i,i+1} J_{i+1}^-, G_{i+1}^- = T_{i,i+1} J_i^+$ via the transmittance $T_{i,i+1}$ of the layer between planes *i* and *i+1*. The transmittance is given by the Beer-Lambert Law $T_{i,i+1} = exp(-\alpha_{i,i+1} t_{i,i+1}/cos\theta_{i,i+1})$, where $t_{i,i+1}, \alpha_{i,i+1}, \theta_{i,i+1}$ are the thickness, absorption constant and propagation angle in the medium between the planes *i* and *i+1,* respectively. The intensity at each interface satisfies $J_i^\pm = (1 - R_i^{s,p}) G_i^\mp + R_i^{s,p} G_i^\pm$, where $R_i^{s,p}(\omega,\theta)$, is the Fresnel reflectance for *s,p* waves. The boundary conditions for pump absorption are the incident sunlight from only the upper side, i.e., $G_0^- = I_{sun}(\omega,\theta), G_N^+ = 0$, where $I_{sun}(\omega,\theta)$ is the solar flux. These linear sets of equations are solved for $J_i^\pm(\omega,\theta)$ to calculate the absorption in each layer, $Abs_{i,i+1}(\omega,\theta) = (1 - T_{i,i+1})(J_i^+ + J_{i+1}^-)$. Normal incidence is assumed for a micro-laser that is pumped with a low solar concentration, i.e.,

$\theta_{sun} = 0$. The frequency-dependent absorption is integrated over the solar spectrum to give the total absorbed photon flux in the sensitizer.

After the excitation profile within the luminescent **sensitizing layer** is known, the modified boundary conditions for the luminescence light absorption in the **gain media** can be defined. Assuming that there is no direct pumping of the gain media, then no light is impinging on the micro-laser from either side, i.e., $G_0^- = 0, G_N^+ = 0$; instead, the light is generated within the sensitizer layer. The worst case approximation would be to assume that the luminescent light is uniformly emitted (see Supplementary section II for details on absorption of luminescence and section III for details on non-uniform emission) in the immediate vicinity of the top surface, which enhances the calculated self-absorption. Therefore, the equations for $G_0^+, J_0^+$ are modified to account for the excitation: $G_0^+ = T_{0,1} J_1^- + luminescense = T_{0,1} J_1^- + \frac{1}{2} I_{sens}(\omega, \theta_1)$, $J_0^+ = (1 - R_0^{s,p}) G_0^- + R_0^{s,p} G_0^+ + luminescense = (1 - R_0^{s,p}) G_0^- + R_0^{s,p} G_0^+ + \frac{1}{2} I_{sens}(\omega, \theta_1)$, where $I_{sens}(\omega, \theta_1)$ stands for frequency weighted emission of the sensitizer with a uniform distribution inside the sensitizer layer; and $0 \leq \theta_1 \leq \frac{\pi}{2}$. Integration on the emission of the sensitizer, $I_{sens}(\omega, \theta_1)$, over the frequency and angles $0 \leq \theta_1 \leq \frac{\pi}{2}$ in the sensitizer (see Supplementary section IV for modeling the complex angle of incidence) corresponds to the absorbed solar photon flux multiplied by the sensitizer quantum efficiency $\eta_s$. To obtain the photon flux absorbed in the gain media, $Abs_{gain}(\omega, \theta_1)$ (per angle, per frequency, per polarization), the modified equations are solved for $J_i^{\pm}(\omega, \theta_1)$. Next, we integrate $Abs_{gain}(\omega, \theta_1)$ over the sensitizer emission frequencies and for $0 \leq \theta_1 \leq \frac{\pi}{2}$. The contributions of the *s* and *p* polarizations are added. The second-order effects of self-absorption and re-emission are also included in the calculation and are modeled as a sum of infinite series of absorption-emission events in the sensitizer, the same as for the LSCs[30] (see the Supplementary section V).

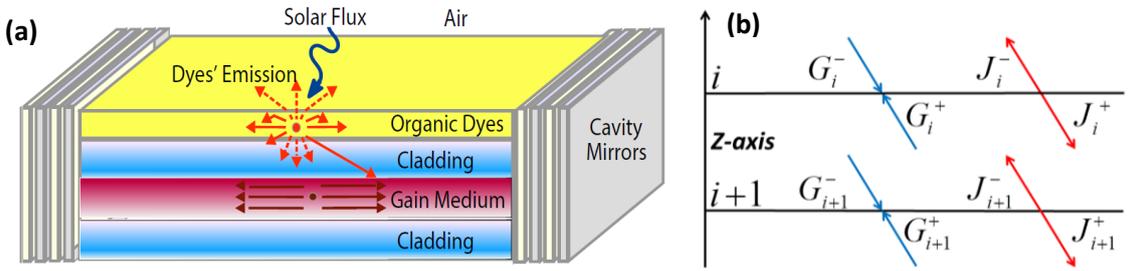

**Figure 1** (**a**) A concept device: The pump light is absorbed by a layer of luminescent dyes and is re-emitted into the waveguide, and a fraction of this luminescence is captured in the structure. The captured photons are absorbed by the gain media or reabsorbed by the sensitizer. (**b**) A schematic of radiation net transfer at interfaces *i,i+1*. Radiation at each interface on either side is modeled as a sum of incoming and outgoing intensities that impinge at a specific angle, polarization and frequency.

As an example for the above analysis, we examine an ideally sensitized 1%at. Nd³⁺:YAG SPLs. Nd³⁺:YAG has main absorption line at 808 nm[32] and lasing wavelength $\lambda_L = 1064$[32-35]. Therefore an ideal sensitizer for Nd³⁺:YAG has its emission centered at 808 nm, unity quantum efficiency $\eta_s = 100\%$ and absorption cutoff at around $\lambda_a = 710$ nm, taking into account typical sensitizer Stokes' shift. The ratio of the sensitizer absorption constants at the absorption band, $\alpha_a$, to its self-absorption constant, $\alpha_s$, assumed to be on the same scale as $\alpha_{Nd^{3+}:YAG}^{808nm}/\alpha_{Nd^{3+}:YAG}^{1064nm}$, where $\alpha_{Nd^{3+}:YAG}^{808nm}, \alpha_{Nd^{3+}:YAG}^{1064nm}$ is the absorption constants of Nd$^{3+}$:YAG at 808 nm and its distributed loss constant at 1064 nm, respectively. Hence, we define $\alpha_a = 10^6\ m^{-1}$, $\alpha_s = 10^3\ m^{-1}$, where the last value is approximately the same as the Nd$^{3+}$:YAG absorption coefficient $\alpha_g = \alpha_{Nd^{3+}:YAG}^{808nm} \sim \alpha_s$. Based on the above estimation the effective pumping regime is reached for cavities much thicker than $\alpha_s/(\alpha_g \alpha_a) \sim 1\ \mu m$. This is in contrast for the required thin cavity supporting low $P_{th}$. The lasing cavity is assumed to be $l$=1 cm in length with an output coupler mirror reflectivity that matches the roundtrip cavity material loss of $R_{oc} = exp(-2l\alpha_{Nd^{3+}:YAG}^{1064nm})$. In this case, half of the total power given off by the Nd$^{3+}$ atoms due to the stimulated emission is coupled out of the laser[36], and the other half is lost owing to material losses. Figure **2a** depicts the energy transfer efficiency between the sensitizer and the gain media. This depiction is accomplished by calculating the fraction of photons absorbed in the gain media normalized to the total emitted photons by the sensitizer, which is calculated for various sensitizer and gain media thicknesses. A thicker sensitizer layer means that a larger fraction of the pump is absorbed, but a smaller fraction of photons reaches the gain media owing to the rise in the self-absorption at the sensitizer. For a sufficiently thick cavity and a sufficiently thin sensitizer layer, i.e., $t_s \alpha_s \ll t_g \alpha_g$, the ratio of absorbed photons in the gain media reaches the value of the captured photons in the waveguiding structure of $\eta_{trap} \approx 80\%$[30], which is limited only by Snell's law, but such a regime in not optimal owing to poor pump absorption. The tradeoff between the absorption and self-absorption leads to a distinct optimum sensitizer thickness when calculating the solar concentration at the lasing threshold (Figure **2b**), which is defined as $C_{th}^{solar} = \left(\frac{N_t V}{\tau_{sp}}\right)/(S \cdot Abs_{gain}^{1sun})$. $N_t = 2 \cdot \alpha_{Nd^{3+}:YAG}^{1064nm}/\sigma_{Nd^{3+}:YAG}^{1064nm}$ is the population inversion of the cavity at threshold, $\sigma_{Nd^{3+}:YAG}^{1064nm} = 8.8 \cdot 10^{-19} cm^2$ is the emission cross-section at the lasing wavelength, and factor two arises from equating the output coupler and cavity material losses. $V$ is the volume of the gain media, $S$ is the surface area of the device, $\tau_{sp} = 230\ \mu sec$ is the Nd$^{3+}$:YAG fluorescence lifetime, and $Abs_{gain}^{1sun}$ is the photon flux absorbed in the gain media, when the sensitizer is pumped by non-concentrated solar radiation. As seen, the optimal sensitizer thickness is defined by the gain media thickness via the absorption tradeoff. Figure **2c** depicts the laser slope efficiency for various thicknesses, and it also shows the optimum value due to the competition in absorption. The slope efficiency is defined as $\eta_{slope} = \eta_{oc}\ \eta_{Nd^{3+}:YAG} Abs_{gain}^{1sun} h\nu_L/P_{sun}$, where $\eta_{oc} = \frac{1-R_{oc}}{1-exp(-2l\alpha_{Nd^{3+}:YAG}^{1064nm})+(1-R_{oc})} = \frac{1}{2}$ is the useful output, i.e., the fraction of the total power that is coupled out of the laser and is adjusted via the output coupler reflectivity[36]. Here, $\eta_{Nd^{3+}:YAG} = 65\%$ is the Nd$^{3+}$:YAG quantum efficiency, $h\nu_L$ is the energy of the photon at the lasing frequency $\nu_L$, and $P_{sun} = 1 kW m^2$ is the solar power per unit area. In contrast to conventional lasers, in which full absorption is desired, here the optimum sensitizer thickness does not correspond to full solar absorption and scales as $t_g \alpha_g$. The ratios $\alpha_a/\alpha_s$ and $\alpha_g/\alpha_s$ define the sensitivity of the optima and strongly influence the fraction of the absorbed photons in the gain medium. Figure **2d** analyzes the optimal conditions for the SPLs, i.e., the sensitizer layer is set at the optimal value for each gain medium thickness. The optimal pump absorption is defined by the optimal sensitizer layer thickness and is shown in magenta (magenta solid line, left magenta axis). The minimal solar concentration at threshold is shown to be less than one sun for a gain media that is thinner than 35 $\mu m$ (blue line, right blue axis). The maximal slope efficiency of SPL (red line, left red axis) is at the optimal absorption value (optimal sensitizer thickness); it increases with the gain media thickness because the thicker gain media allows a thicker sensitizer and consequently a higher pump absorption. Thus, ideally sensitized $Nd^{+3}$:YAG-based SPLs could operate at a non-concentrated solar pump with a 5% slope efficiency. The maximal slope efficiency saturates as the optimal sensitizer thickness reaches full pump absorption.

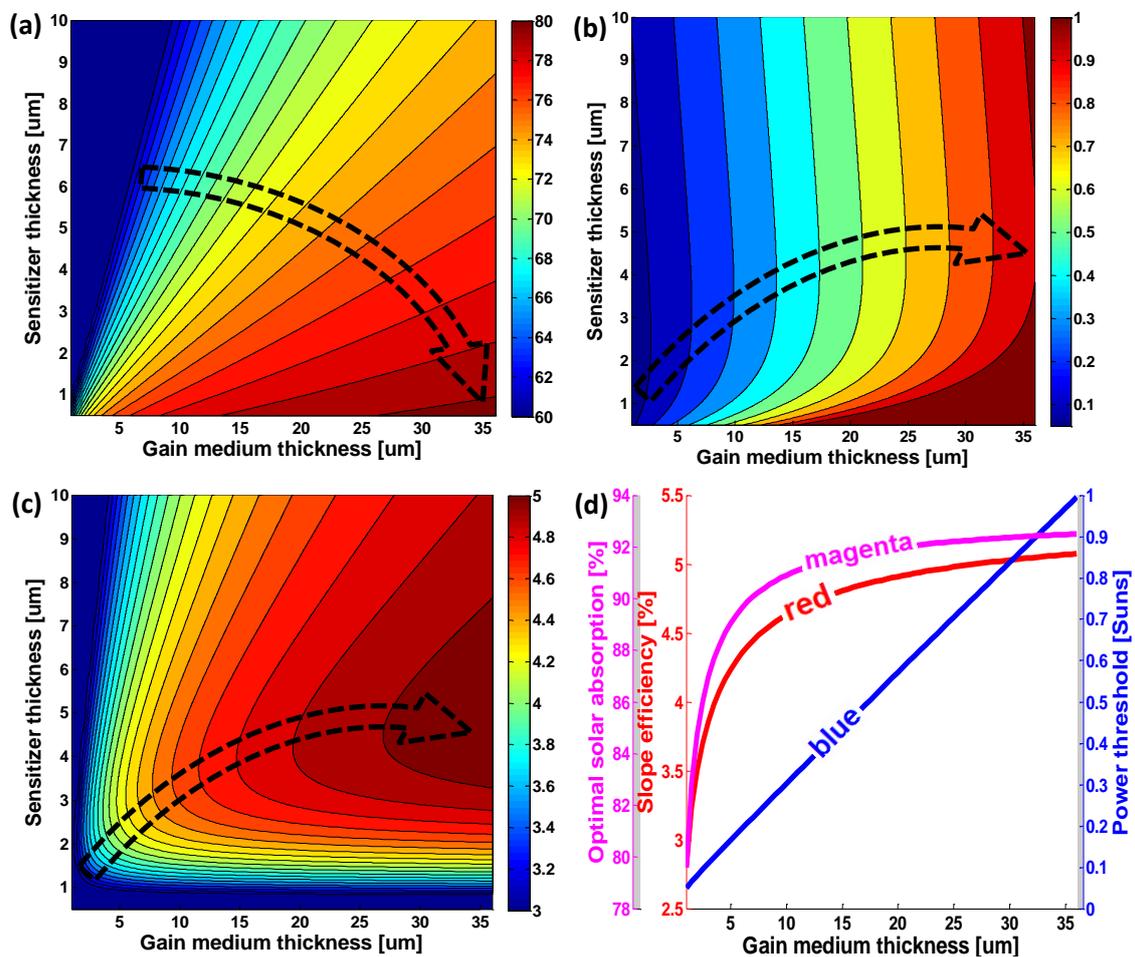

**Figure 2** (**a**) The energy transfer between the sensitizer and the gain media is depicted as a fraction of the emitted photoluminescent photons that are absorbed by the gain medium. (**b**) Required solar concentration at the lasing threshold. (**c**) Slope efficiency when the output coupler loss is matched with material losses. Black dotted arrows point to the direction of growing values. (**d**) Slope efficiency (red line, left red axis), solar concentration at threshold (blue line, right blue axis) and optimal absorption (magenta line, left magenta axis) per cavity thickness with the optimal sensitizer layer.

Notably, in the thin planar waveguide, the lasing mode tail overlaps with the high absorption region of the sensitizer, which significantly affects the resonator Q-factor. Therefore, it is constructive to induce a spatial separation between the sensitizer and the gain medium layers. In contrast to near-field sensitization, radiative energy transfer allows avoiding this negative effect by introducing lossless cladding with an intermediate refractive index between the sensitizer and gain medium (as shown in Figure **1a**), which effectively confines the lasing mode in the low loss region and increases the Q-factor to the value of the unperturbed cavity. The results presented in Figure **2** are relevant for this case as they assume lossless cladding (see SUPPLEMENTARY section VI for the discussion).

As experimental validation, we apply our theory to an organic sensitizer and $Nd^{3+}$:YAG planar cavity. The sensitizer is composed of a combination of dyes $AlQ_3$:DCJTB(2%):Pt(TPBP)(4%)[28] that harvest solar radiation between 350 nm and 650 nm. Figure **3a** shows the absorption coefficient of this sensitizer (red line, left red logarithmic axis), overlaid with solar photon flux (magenta line, right magenta axis). Nonradiative (near field) energy transfer[15–18] from the $AlQ_3$ to the DCJTB continuing to the Pt(TPBP) with close to unity efficiency allows to reduce the concentration of the emitting dye - Pt(TPBP), which results in its absorption constant being orders of magnitude higher than for $Nd^{3+}$:YAG in the visible spectrum (Fig. **3a**, blue line, left blue axis) while maintaining low self-absorption, below the $Nd^{3+}$:YAG absorption (Fig. **3b**, red solid line, right logarithmic axis). The Pt(TPBP) has an emission peak at 780 nm with a full-width half-maximum of 50 nm (Fig. **3b**, magenta line, left axis). This emission overlaps with the $Nd^{3+}$:YAG absorption lines, as shown in the blue solid line in Fig. **3b** (right logarithmic axis). The value of the Pt(TPBP) self-absorption coefficient that is assumed in our simulations is the maximal value of the experimental data and approximated self-absorption (Fig. **3b**, red solid and gray dashed lines, respectively, right logarithmic axis). We have grown a 3.2-$\mu m$-thick sensitizer layer via thermal deposition on a glass slide to characterize the absorption and external quantum efficiency (Figure **3c**, red and blue lines, respectively), under incoherent continuous-wave excitation at various wavelengths. To measure the pumping efficiency, which is the ratio of the absorbed pump photons to the photons emitted by the $Nd^{3+}$, the same sensitizing layer was grown on a 750-$\mu m$-thick $Nd^{3+}$:YAG slab. As shown in Fig. **3c**, magenta solid line, approximately 27% of the absorbed photons are transferred to $Nd^{3+}$ emission. This value is due to the quantum efficiency of $Nd^{3+}$:YAG, $\eta_{Nd^{3+}:YAG} = 65\%$[34], the sensitizer quantum efficiency $\eta_s \approx 50\%$ and the trapping efficiency $\eta_{trap} \approx 80\%$[30]. Comparing these results to the theoretical model, we note that for the 750-$\mu m$-thick waveguide and 3.2-$\mu m$-thick sensitizer, the condition $t_s \alpha_s \ll t_g \alpha_g$ is satisfied since $t_g \alpha_g / t_s \alpha_s > 200$. For such a case, the energy transfer efficiency between the sensitizer and the gain media is at the maximal theoretical value of $\eta_{trap} \approx 80\%$, and therefore, the overall photon transfer efficiency is $\eta_s \eta_{trap} \eta_{Nd^{3+}:YAG} \approx 27\%$. Using the theoretical model with the experimental data on the quantum efficiency (Fig. **3c**, blue line) results in the predicted energy transfer per wavelength to the $Nd^{3+}$ emission. This prediction is depicted by the green dashed line in Figure **3c** and shows conformity with the measured values (Fig. **3c**, magenta solid line).

Based on our simulations and experimental data, such a sensitizer makes it possible to construct SPLs that operate under non-concentrated solar illumination. As shown in Figure **3d,** the solar threshold (blue line, right blue axis) reaches a non-concentrated condition at a cavity thickness of 3 $\mu m$ (blue line below 1 sun). Unfortunately such cavities are not readily available and the available sensitizers saturate at intensities of $P_{sun}$ and above (see SUPPLEMENTARY section VII) and an actual device yet to be demonstrated. Moreover, in this case, the slope efficiency is only 0.53% (Fig. **4d**, red line, left red axis). Such a low slope efficiency is a result of few factors: **i.** the self-absorption assumed in our simulations (gray dashed line in Fig. **4b**), **ii.** non-unity quantum efficiency of the gain media and sensitizer (blue solid line in Fig. **4c**), and **iii.** a mismatch between the sensitizer emission and peak $Nd^{3+}$ absorption line (magenta and blue solid lines in Figure **4b**), which results in a disadvantageous ratio of sensitizer self-absorption and $Nd^{3+}$:YAG absorption at the sensitizer emission wavelength. These losses affect the absorption tradeoff and set the optimal sensitizer absorption at the very low value of only 38% (Fig. **4d**, magenta line, left magenta axis), which becomes the main limitation of the overall efficiency.

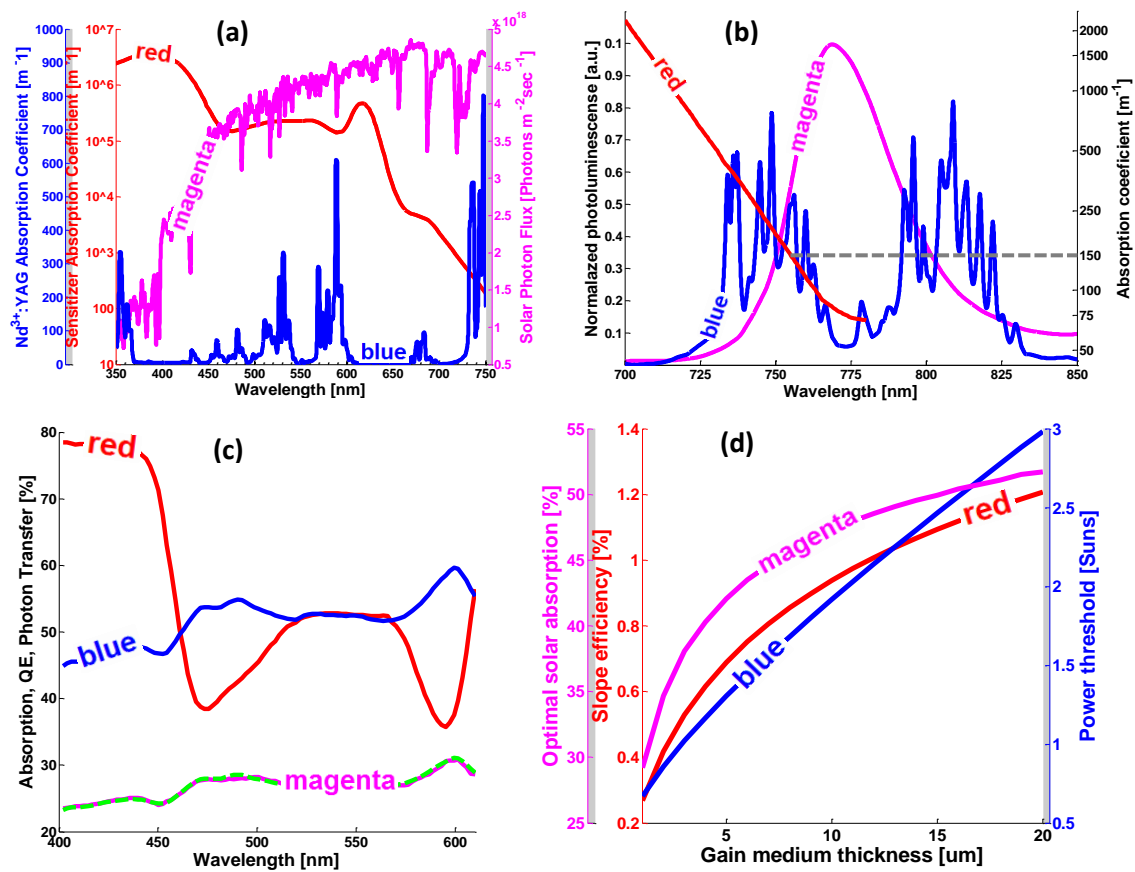

**Figure 3** (**a**) $Nd^{3+}$:YAG (blue line, left blue axis) and organic sensitizer (red line, left red logarithmic axis) absorption constant in the visible spectral range overlaid with the solar flux (magenta line, right magenta axis). (**b**) Pt(TPBP) luminescence spectrum (magenta line, left axis) overlaps the $Nd^{3+}$:YAG absorption coefficient (blue line, right logarithmic axis). Red solid line and gray dashed line depict the sensitizer measured and the approximated self-absorption constants, respectively (right logarithmic axis). (**c**) Experimentally measured absorption of the 3.2-µm-thick $AlQ_3$:DCJTB(2%):Pt(TPBP)(4%) layer (red solid line) and its external quantum efficiency (blue solid line). The magenta solid and green dashed lines present the measured and predicted ratios between the $Nd^{3+}$ emission rate and the initially absorbed photon rate for the 3.2-µm-thick sensitizer deposited on the 750-µm-thick $Nd^{3+}$:YAG slab cavity. (**d**) Predicted slope efficiency (red line, left red axis), solar concentration at threshold (blue line, blue right axis) and optimal absorption (magenta line, left magenta axis) per cavity thickness with optimal sensitizer thickness.

**In conclusion**, we present a generic theoretical framework for designing a broadband pump schema for micro-lasers and SPLs based on the net radiation method. The formalism reveals that the interplay between the sensitizer self-absorption and cavity absorption defines the optimal value for the pump absorption. This finding is in sharp contrast to conventional lasers in which full absorption is desired. The presented theoretical approach is generic and modular, which allows individual explicit optimization of each component of the broadband-pumped high-Q lasers, providing the guidelines for optimal configuration. Beyond on-chip micro-lasers and SPLs, this method can be useful for other illuminating devices such as LSCs.

**METHODS**

The 750-μm-thick $Nd^{3+}$:YAG slab waveguide was formed through conventional polishing techniques from commercial laser rods with a diameter of 3 mm and length of 43 mm, with λ=1064 nm antireflection coatings on both faces. A 3.2-μm-thick film of $AlQ_3$:DCJTB(2%):Pt(TPBP)(4%) was thermally evaporated on the $Nd^{3+}$:YAG slab, and layers of various thicknesses were thermally deposited on microscopic glass slides for preliminary experiments. The $AlQ_3$:DCJTB(2%):Pt(TPBP)(4%), $Nd^{3+}$:YAG and $Nd^{3+}$:YAG with $AlQ_3$:DCJTB(2%):Pt(TPBP)(4%) films were excited with monochromatic light. The monochromatic light was generated using a tungsten-bulb monochromator with a spectral resolution of 3.5 nm full-width at half maximum and a mechanical chopper. The sample photoluminescence was measured with a lock-in amplifier and a spectrally calibrated photodetector, and long wavelength filters with a cutoff at λ=850 nm were used to discriminate between $Nd^{3+}$:YAG and $AlQ_3$:DCJTB(2%):Pt(TPBP)(4%) emission in the integrating sphere. An Acton spectrometer with a grating of 1000 grooves per mm and a calibrated Newport photodetector and neutral density filter were used in all of the spectral measurements. All of the simulations for solving the net radiation method equations were performed on a desktop computer with 32 GB RAM and are detailed in the supplementary materials.

**RFERENCES**

1. Vollmer, F. *et al.* Protein detection by optical shift of a resonant microcavity. *Appl. Phys. Lett.* **80,** 4057–4059 (2002).

2. Serpengüzel, A., Griffel, G. & Arnold, S. Excitation of resonances of microspheres on an optical fiber. *Opt. Lett.* **20,** 654 (1995).

3. Spillane, S. M., Kippenberg, T. J. & Vahala, K. J. Ultralow-threshold Raman laser using a spherical dielectric microcavity. *Nature* **415,** 621–623 (2002).

4. Treussart, F. *et al.* Evidence for intrinsic Kerr bistability of high-Q microsphere resonators in superfluid helium. *Eur. Phys. J. D* **1,** 235 (1998).

5. Little, B. E. *et al.* Very high-order microring resonator filters for WDM applications. *IEEE Photonics Technol. Lett.* **16,** 2263–2265 (2004).

6. Ferrera, M. *et al.* Low-power continuous-wave nonlinear optics in doped silica glass integrated waveguide structures. *Nat. Photonics* **2,** 737–740 (2008).

7. Kiss, Z. J., Lewis, H. R. & Jr, R. C. D. SUN PUMPED CONTINUOUS OPTICAL MASER. *Appl. Phys. Lett.* **2,** 93–94 (1963).

8. Liang, D. & Almeida, J. Highly efficient solar-pumped Nd:YAG laser. *Opt. Express* **19,** 26399–26405 (2011).

9. Yabe, T. *et al.* High-efficiency and economical solar-energy-pumped laser with Fresnel lens and chromium codoped laser medium. *Appl. Phys. Lett.* **90,** 261120 (2007).

10. Dinh, T. H., Ohkubo, T., Yabe, T. & Kuboyama, H. 120 watt continuous wave solar-pumped laser with a liquid light-guide lens and an Nd:YAG rod. *Opt. Lett.* **37,** 2670–2672 (2012).

11. Yagi, H., Yanagitani, T., Yoshida, H., Nakatsuka, M. & Ueda, K. Highly Efficient Flashlamp-Pumped $Cr^{3+}$ and $Nd^{3+}$ Codoped $Y_3Al_5O_{12}$ Ceramic Laser. *Jpn. J. Appl. Phys.* **45,** 133–135 (2006).

12. Dong, J. *et al.* Composite Yb:YAG/$Cr^{4+}$:YAG ceramics picosecond microchip lasers. *Opt. Express* **15,** 14516 (2007).

13. Yabe, T. *et al.* 100 W-class solar pumped laser for sustainable magnesium-hydrogen energy cycle. *J. Appl. Phys.* **104,** 83104 (2008).

14. Mizuno, S., Ito, H., Hasegawa, K., Suzuki, T. & Ohishi, Y. Laser emission from a solar-pumped fiber. *Opt. Express* **20,** 5891–5895 (2012).

15. Förster, T. Transfer Mechanisms of Electronic Excitation Energy. *Radiat. Res. Suppl.* **2,** 326–339 (1960).

16. Lakowicz, J. R. *Principles of Fluorescence Spectroscopy*. (Springer US, 2006).

17. Masters, B. R. Paths to Förster's resonance energy transfer (FRET) theory. *Eur. Phys. J. H* **39,** 87–139 (2013).

18. Andrews, D. L. in *Tutorials in Complex Photonic Media* (eds. Noginov, M. A., McCall, M. W., Dewar, G. & Zheludev, N. I.) 439–478 (SPIE Press, 2009).

19. Berggren, M., Dodabalapur, A., Slusher, R. E. & Bao, Z. Light amplification in organic thin films using cascade energy transfer. *Nature* **389,** 466–469 (1997).

20. Zhao, X. K., Zhao, Q. W. & Shen, X. F. Lasing Action in Dual-Doped Organic Microcavity with Cascade Energy Transfer. *Adv. Mater. Res.* **503–504,** 1125–1128 (2012).

21. Shopova, S. I. *et al.* Opto-fluidic ring resonator lasers based on highly efficient resonant energy transfer. *Opt. Express* **15,** 12735 (2007).

22. Rotschild, C. *et al.* Cascaded Energy Transfer for Efficient Broad-Band Pumping of High-Quality, Micro-Lasers. *Adv. Mater.* **23,** 3057–3060 (2011).

23. Reusswig, P. D. *et al.* A path to practical Solar Pumped Lasers via Radiative Energy Transfer. *Sci. Rep.* **5,** 14758 (2015).

24. Wang, L., Jacques, S. L. & Zheng, L. MCML—Monte Carlo modeling of light transport in multi-layered tissues. *Comput. Methods Programs Biomed.* **47,** 131–146 (1995).

25. Siegel, R. *Net radiation method for enclosure systems involving partially transparent walls*. (1973).

26. Grivas, C. Optically pumped planar waveguide lasers, Part I: Fundamentals and fabrication techniques. *Prog. Quantum Electron.* **35,** 159–239 (2011).

27. Grivas, C. Optically pumped planar waveguide lasers: Part II: Gain media, laser systems, and applications. *Prog. Quantum Electron.* **45–46,** 3–160 (2016).

28. Currie, M. J., Mapel, J. K., Heidel, T. D., Goffri, S. & Baldo, M. A. High-Efficiency Organic Solar Concentrators for Photovoltaics. *Science* **321,** 226–228 (2008).

29. Weber, W. H. & Lambe, J. Luminescent greenhouse collector for solar radiation. *Appl. Opt.* **15,** 2299–2300 (1976).

30. Batchelder, J. S., Zewai, A. H. & Cole, T. Luminescent solar concentrators. 1: Theory of operation and techniques for performance evaluation. *Appl. Opt.* **18,** 3090–3110 (1979).

31. Meinardi, F. *et al.* Large-area luminescent solar concentrators based on `Stokes-shift-engineered' nanocrystals in a mass-polymerized PMMA matrix. *Nat. Photonics* **8,** 392–399 (2014).

32. Semwal, K. & Bhatt, S. C. Study of $Nd^{3+}$ ion as a Dopant in YAG and Glass Laser. *Int. J. Phys. Int. J. Phys.* **1,** 15–21 (2013).

33. Birnbaum, M. & Klein, C. F. Stimulated emission cross section at 1.061 μm in Nd:YAG. *J. Appl. Phys.* **44,** 2928–2930 (1973).

34. Singh, S., Smith, R. G. & Van Uitert, L. G. Stimulated-emission cross section and fluorescent quantum efficiency of $Nd^{3+}$ in yttrium aluminum garnet at room temperature. *Phys. Rev. B* **10,** 2566–2572 (1974).

35. Kushida, T., Marcos, H. M. & Geusic, J. E. Laser Transition Cross Section and Fluorescence Branching Ratio for $Nd^{3+}$ in Yttrium Aluminum Garnet. *Phys. Rev.* **167,** 289–291 (1968).

36. Yariv, A. *Photonics : optical electronics in modern communications /*. (Oxford University Press, 2007).

## ACKNOWLEDGEMENTS

This report was partially supported by the Russell Berrie Nanotechnology Institute (RBNI) and the Grand Technion Energy Program (GTEP) and is part of The Leona M. and Harry B. Helmsley Charitable Trust reports on the Alternative Energy series of the Technion and the Weizmann Institute of Science. This report was also partially support by the Israel Strategic Alternative Energy Foundation (I-SAEF). The authors would also like to acknowledge the partial support provided by the Focal Technology Area on Nanophotonics for Detection. Prof. C. Rotschild would like to thank the Marie Curie European Reintegration Grant for its support.

## AUTHOR CONTRIBUTIONS

S.N. developed the theoretical models and wrote the paper, P.D.R. performed the sensitizer deposition and characterization experiments, C.R. and M.A.B. conceived the original idea and contributed to the writing of the paper.

## ADDITIONAL INFORMATION

**Competing financial interests**: The authors declare no competing financial interests.

## FIGURE LEGENDS

**Figure 1** (**a**) A concept device: The pump light is absorbed by a layer of luminescent dyes and is re-emitted into the waveguide, and a fraction of this luminescence is captured in the structure.

The captured photons are absorbed by the gain media or reabsorbed by the sensitizer. (**b**) A schematic of radiation net transfer at interfaces *i,i+1*. Radiation at each interface on either side is modeled as a sum of incoming and outgoing intensities that impinge at a specific angle, polarization and frequency.

**Figure 2** (**a**) The energy transfer between the sensitizer and the gain media is depicted as a fraction of the emitted photoluminescent photons that are absorbed by the gain medium. (**b**) Required solar concentration at the lasing threshold. (**c**) Slope efficiency when the output coupler loss is matched with material losses. Black dotted arrows point to the direction of growing values. (**d**) Slope efficiency (red line, left red axis), solar concentration at threshold (blue line, right blue axis) and optimal absorption (magenta line, left magenta axis) per cavity thickness with the optimal sensitizer layer.

**Figure 3** (**a**) $Nd^{3+}$:YAG (blue line, left blue axis) and organic sensitizer (red line, left red logarithmic axis) absorption constant in the visible spectral range overlaid with the solar flux (magenta line, right magenta axis). (**b**) Pt(TPBP) luminescence spectrum (magenta line, left axis) overlaps the $Nd^{3+}$:YAG absorption coefficient (blue line, right logarithmic axis). Red solid line and gray dashed line depict the sensitizer measured and the approximated self-absorption constants, respectively (right logarithmic axis). (**c**) Experimentally measured absorption of the 3.2-μm-thick $AlQ_3$:DCJTB(2%):Pt(TPBP)(4%) layer (red solid line) and its external quantum efficiency (blue solid line). The magenta solid and green dashed lines present the measured and predicted ratios between the $Nd^{3+}$ emission rate and the initially absorbed photon rate for the 3.2-μm-thick sensitizer deposited on the 750-μm-thick $Nd^{3+}$:YAG slab cavity. (**d**) Predicted slope efficiency (red line, left red axis), solar concentration at threshold (blue line, blue right axis) and optimal absorption (magenta line, left magenta axis) per cavity thickness with optimal sensitizer thickness.

# SUPPLEMENTARY: Designing a Broadband Pump for High-Quality Micro-Lasers via Modified Net Radiation Method

Sergey Nechayev[1], Philip D. Reusswig[2], Marc. A. Baldo[2], Carmel Rotschild[1*]

[1]Department of Mechanical Engineering and Russell Berrie Nanotechnology Institute,
Technion-Israel Institute of Technology, Haifa 32000, Israel

[2]Department of Electrical Engineering and Computer Science, Massachusetts Institute of
Technology, 77 Massachusetts Avenue, Cambridge, MA 02139, USA

*Corresponding authors: carmelr@technion.ac.il

**I-Net radiation method for solar light absorption**

We utilize the net radiation method[1], which is a convenient tool for addressing incoherent light absorption in thin films, to calculate the solar light absorption in a stratified medium. The incoherent light approximation remains valid as long as the source coherence length is smaller than the optical path in the sensitizer layer, $l_c^{sun} \leq Re(n_s)\frac{t_s}{cos(\theta)}$, where $l_c^{sun}$ is the solar light coherence length, $t_s$ is the thickness of the sensitizer length, θ is the angle of propagation of the light in the sensitizer, and $n_s$ is its refractive index. This condition holds true for layers that are thicker than 350 nm because $l_c^{sun} \sim 0.6 \mu m, Re(n_s) = 1.7$.

All of the calculations are presented for 2-layer structures for clarity, and they are easily extended to any number of layers by utilizing the more general equations that are presented in the manuscript. In the manuscript, the simulations are performed for 4-layer devices with non-absorptive cladding (as depicted in Fig. **1a** in the manuscript).

Consider the structure in Figure **S1** and the media enclosed between planes *i* and *i+1*. In the net radiation method, the optical field intensity is defined as the sum of the forward and backward propagating intensity waves. At the *i*th plane, the outgoing and incoming intensities are designated as $J_i^{\pm}(\omega, \theta)$ and $G_i^{\pm}(\omega, \theta)$, respectively, where $\omega, \theta$ stand for the angular frequency and angle of incidence, respectively. Planar systems have axial symmetry, and therefore, $\theta$ measured in any layer defines the angle in all of the other layers, according to Snell's law. Our sign convention is that " + " defines intensity components that are situated in the medium above the interface, and " − " is for the medium below the interface. In addition, in each medium, the incoming and outgoing intensity components are connected by the equations $G_i^+ = T_{i,i+1} J_{i+1}^-, G_{i+1}^- = T_{i,i+1} J_i^+$ via the transmittance $T_{i,i+1}$ of the layer between planes *i* and *i+1*.

First, we start with the case of solar light absorption in the sensitizer layer. In this case, the boundary conditions correspond to the solar flux at normal incidence from the positive direction of the z-axis. $G_0^- = I_{sun}(\omega, \theta), G_2^+ = 0$, where $I_{sun}(\omega, \theta)$ is the solar flux per frequency. First, we write the equations that connect the inbound intensities with the outbound intensities through the specular reflectance coefficients.

$J_0^- = R_s G_0^- + (1 - R_s) G_0^+ = R_s I_{sun}(\omega) + (1 - R_s) G_0^+$

$J_0^+ = (1 - R_s) G_0^- + R_s G_0^+ = (1 - R_s) I_{sun}(\omega) + R_s G_0^+$

$J_1^- = R_{sg} G_1^- + (1 - R_{sg}) G_1^+$

$J_1^+ = (1 - R_{sg}) G_1^- + R_{sg} G_1^+$

$J_2^- = R_g G_2^- + (1 - R_g) G_2^+ = R_g G_2^-$

$J_2^+ = (1 - R_g) G_2^- + R_g G_2^+ = (1 - R_g) G_2^-$

$R_s(\theta, \lambda), R_g(\theta, \lambda), R_{sg}(\theta, \lambda)$, are the polarization-dependent Fresnel's reflectance coefficients at the air-sensitizer, air-gain media and sensitizer-gain media interfaces, respectively and $\lambda$ is a free space wavelength. Next, the relations between the inbound and outbound components through the transmittance are used to eliminate the outbound components:

$$G_0^+ = T_s J_1^- \;;\; G_1^- = T_s J_0^+$$
$$G_1^+ = T_g J_2^- = J_2^- \;;\; G_2^- = T_g J_1^+ = J_1^+$$

Here, the transmittance of the gain medium $T_g(\lambda, \theta_g) = 1$, i.e., we neglect the direct pump absorption in the gain media owing to its expected thickness (on the order of several microns) and low gain medium absorption coefficient $\alpha_g(\lambda)$ in the visible spectral range. The sensitizer absorption follows an exponential Beer-Lambert Law at normal incidence, and thus, sensitizer transmittance $T_s = e^{-\alpha_s(\lambda) t_s}$, where $\alpha_s(\lambda)$ is the absorption constant of the sensitizer. We rewrite the equations for the $J_i^{\pm}$ components:

$$J_0^- = R_s I_{sun}(\theta) + (1 - R_s) T_s J_1^-$$
$$J_0^+ = (1 - R_s) I_{sun}(\theta) + R_s T_s J_1^-$$
$$J_1^- = R_{sg} T_s J_0^+ + (1 - R_{sg}) J_2^-$$
$$J_1^+ = (1 - R_{sg}) T_s J_0^+ + R_{sg} J_2^-$$
$$J_2^- = R_g J_1^+$$
$$J_2^+ = (1 - R_g) J_1^+$$

Because we are interested in only 4 of the 6 components, we can rewrite them in a convenient linear equation form:

$$\mathbf{y} = \mathbf{A} \cdot \mathbf{x} + \mathbf{b}$$

$$\begin{pmatrix} J_0^+ \\ J_1^- \\ J_1^+ \\ J_2^- \end{pmatrix} = \begin{pmatrix} 0 & R_s T_s & 0 & 0 \\ R_{sg} T_s & 0 & 0 & (1 - R_{sg}) \\ (1 - R_{sg}) T_s & 0 & 0 & R_{sg} \\ 0 & 0 & R_g & 0 \end{pmatrix} \begin{pmatrix} J_0^+ \\ J_1^- \\ J_1^+ \\ J_2^- \end{pmatrix} + \begin{pmatrix} (1 - R_s) I_{sun}(\theta) \\ 0 \\ 0 \\ 0 \end{pmatrix}$$

The resulting equations are solved to obtain the absorbed photon flux in the sensitizer $Abs_s(\lambda, \theta, t_s, t_g)$ per angle of incidence per wavelength. Note that $\theta = 0$ for low solar concentrations, and the gain medium thickness $t_g$ is irrelevant because $T_g(\lambda, \theta_g) = 1$ in the visible spectral range. The resulting absorbed solar photon flux in the sensitizer per wavelength when the sensitizer is pumped by non-concentrated solar illumination is:

$$Abs_s^{sol}(\lambda, t_s) = (1 - T_s)(J_0^+ + J_1^-) = I_{sun}(1 - R_s)(1 - T_s) F_{sun}$$

$$F_{sun} = \frac{1 + R_{sg} T_s - R_{sg} R_g + R_g T_s - 2 R_{sg} R_g^2 T_s}{1 - R_{sg} R_g - R_{sg} R_s R_g T^2{}_s - R_s R_g T^2{}_s + 2 R_{sg} R_s R_g T^2{}_s}$$

Because $R_{sg} \approx R_s \approx R_g \approx 0$ and the polarization is degenerate, we can approximate $F_{sun}$ as

$$F_{sun} \approx 1 + (R_{sg} + R_g)T_s \approx 1$$

$$Abs_s^{sol}(\lambda, t_s) \approx I_{sun}(\lambda)(1 - R_s)(1 - T_s) \approx I_{sun}(\lambda)(1 - R_s)\left(1 - e^{-\alpha_s(\lambda)t_s}\right)$$

In other words, for the normal incidence, the non-reflected part of the solar radiation is absorbed according to the Beer-Lambert Law with the sensitizer absorption constant $\alpha_s(\lambda)$.

To obtain the total absorption in terms of the photon flux, $Abs_s^{sol}(\lambda, t_s)$ is integrated over the solar spectrum with respect to the wavelength.

$$Abs_s^{sol}(t_s) = \int Abs_s^{sol}(\lambda, t_s)d\lambda$$

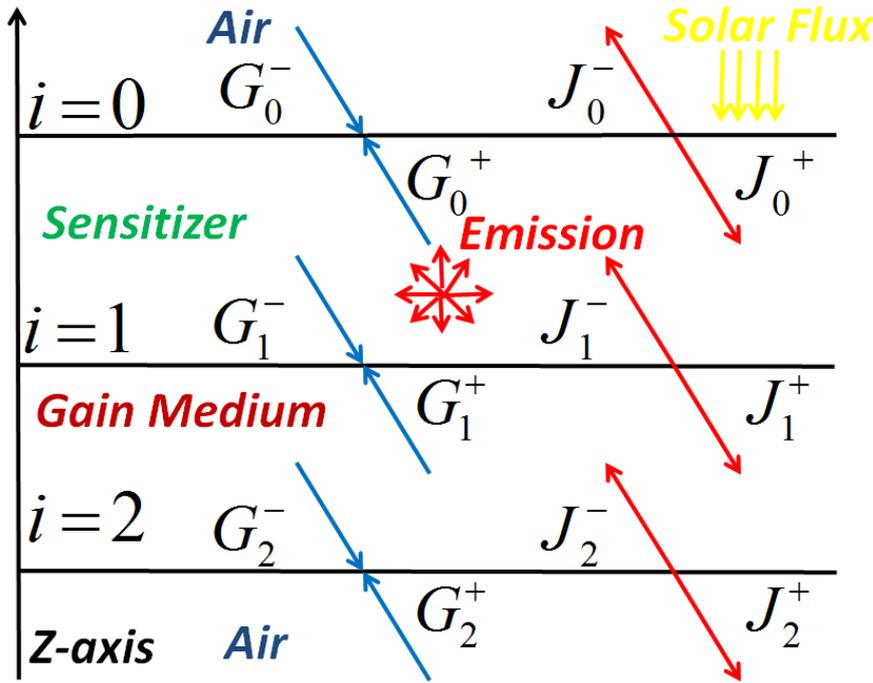

**Figure 1** A schematic of the radiation net transfer at the interfaces between the different layers. The radiation at each interface on either side is modeled as the sum of the incoming and outgoing intensities impinging at a specific angle, polarization and frequency. The former are connected through Fresnel reflectance, transmittance and Snell's law. The boundary conditions are defined by the nature of the excitation. The $J_i^{\pm}, G_i^{\pm}$ values are obtained as a solution of the linear equation system from which the absorption in each layer can be deduced after integration over all of the angles and frequencies for the *s,p* polarizations separately.

**II-Modified net radiation method for the absorption of the luminescence in the gain medium**

After the excitation profile within the luminescent sensitizing layer was determined, the absorption of the luminescence in the gain media and self-absorption of the sensitizer can be defined. Assuming that there is no direct pumping of the gain media, no light is impinging on the structure from either side, i.e., $G_0^- = 0, G_2^+ = 0$; instead, the light is generated within the sensitizer layer. The sensitizer emission is then modeled as uniform non-polarized over angle of incidence in the sensitizer $\theta_s$ in the range $0 \leq \theta_s \leq \frac{\pi}{2}$ with two normalization conditions (see the following chapters for a detailed explanation of $\theta_s$). (**a**) The integration on the emission of the sensitizer, $I_{sens}(\omega, \theta_s)$, over the frequency, polarization and angles $0 \leq \theta_s \leq \frac{\pi}{2}$ in the sensitizer must correspond to the total absorbed solar photon flux multiplied by the sensitizer quantum efficiency $\eta_s$. (**b**) The line-shape of $I_{sens}(\omega, \theta_s)$ follows the sensitizer emission. The worst case approximation would be to assume that the luminescent light is uniformly emitted in the immediate vicinity of the top surface, which accounts for maximal self-absorption (the uniform emission is not a limitation of the model, and non-uniform emission can be easily accounted for by forcing the dependence of $I_{sens}(\omega, \theta_s)$ on $\theta_s$ as explained in the following section). In this case, the equations for $G_0^+, J_0^+$ are modified:

$$G_0^+ = T_s J_1^- + luminescense = T_s J_1^- + \frac{1}{2} I_{sens}(\omega, \theta_s),$$

$$J_0^+ = (1 - R_s^{s,p}) G_0^- + R_s^{s,p} G_0^+ + luminescense = (1 - R_s^{s,p}) G_0^- + R_s^{s,p} G_0^+ + \frac{1}{2} I_{sens}(\omega, \theta_s)$$

where $R_s^{s,p}$ is the polarization-dependent reflectance coefficient at the air-sensitizer interface. The transmittance in each layer follows the Beer-Lambert Law with a propagation distance that depends on the angle of propagation in that medium: $T_s = \exp\left(-\alpha_s \frac{t_s}{\cos\theta_s}\right), T_g = \exp\left(-\alpha_g \frac{t_g}{\cos\theta_g}\right)$. Here $\theta_s, \theta_g$ are the real-valued modeled angles of incidence in the sensitizer and gain media layers, respectively, relative to the optical axis.

The resulting equations are

$$J_0^- = R_s G_0^- + (1 - R_s) G_0^+$$
$$J_0^+ = (1 - R_s) G_0^- + R_s G_0^+ + luminescence$$
$$J_1^- = R_{sg} G_1^- + (1 - R_{sg}) G_1^+$$
$$J_1^+ = (1 - R_{sg}) G_1^- + R_{sg} G_1^+$$
$$J_2^- = R_g G_2^- + (1 - R_g) G_2^+$$
$$J_2^+ = (1 - R_g) G_2^- + R_g G_2^+$$

Substituting all of the above for $G_0^+, J_0^+$ and $G_0^- = 0, G_2^+ = 0$ results in

$$J_0^- = (1 - R_s)\left(T_s J_1^- + \frac{1}{2} I_{sens}\right)$$

$$J_0^+ = R_s \left(T_s J_1^- + \frac{1}{2} I_{sens}\right) + \frac{1}{2} I_{sens}$$

$$J_1^- = R_{sg} T_s J_0^+ + (1 - R_{sg}) J_2^- T_g$$

$$J_1^+ = (1 - R_{sg}) T_s J_0^+ + R_{sg} J_2^- T_g$$

$$J_2^- = R_g J_1^+ T_g$$

$$J_2^+ = (1 - R_g) J_1^+ T_g$$

Rewriting 4 of the above 6 equations in the matrix form $\boldsymbol{y} = \boldsymbol{A} \cdot \boldsymbol{x} + \boldsymbol{b}$, we obtain

$$\begin{pmatrix} J_0^+ \\ J_1^- \\ J_1^+ \\ J_2^- \end{pmatrix} = \begin{pmatrix} 0 & R_s T_s & 0 & 0 \\ R_{sg} T_s & 0 & 0 & (1-R_{sg})T_g \\ (1-R_{sg})T_s & 0 & 0 & R_{sg} T_g \\ 0 & 0 & R_g T_g & 0 \end{pmatrix} \begin{pmatrix} J_0^+ \\ J_1^- \\ J_1^+ \\ J_2^- \end{pmatrix} + \begin{pmatrix} \frac{I_{sens}}{2}(1+R_s) \\ 0 \\ 0 \\ 0 \end{pmatrix}$$

$$Abs_s^{1sun}(\lambda, \theta, t_s, t_g) - = (1-T_s)(J_0^+ + J_1^-) = \frac{I_{sens}}{2}(1+R_s)(1-T_s)F_s(\theta)$$

$$Abs_g^{1sun}(\lambda, \theta, t_s, t_g) - = (1-T_g)(J_1^+ + J_2^-) = \frac{I_{sens}}{2}(1+R_s)(1-T_g)F_g(\theta)$$

Here $Abs_s^{1sun}(\lambda, \theta, t_s, t_g)$ is the self-absorption of the sensitizer, i.e., absorption of the sensitizer emission in the sensitizer layer itself, when the sensitizer is subject to non-concentrated solar irradiation. $Abs_g^{1sun}(\lambda, \theta, t_s, t_g)$ is the absorption of the sensitizer emission in the gain medium, when the sensitizer is subject to non-concentrated solar irradiation.

$$F_s(\theta) = \frac{1 - R_{sg}R_g T^2{}_g + T_s(R_{sg} + R_g T^2{}_g - 2R_{sg}R_g T^2{}_g)}{1 + 2R_{sg}R_s R_g T^2{}_s T^2{}_g - R_{sg}R_s T^2{}_s - R_{sg}R_g T^2{}_g - R_s R_g T^2{}_s T^2{}_g}$$

$$F_g(\theta) = \frac{(1-R_{sg})T_s(R_g T_g + 1)}{1 + 2R_{sg}R_s R_g T^2{}_s T^2{}_g - R_{sg}R_s T^2{}_s - R_{sg}R_g T^2{}_g - R_s R_g T^2{}_s T^2{}_g}$$

To gain intuition, we next examine the trapped light at angles above the critical angle: In this case $R_s = R_g = 1, R_{sg} \approx 0$ and:

$$F_s(\theta) \approx \frac{1 + T_s T_g^2}{1 - T_s^2 T_g^2} \quad F_g(\theta) \approx \frac{T_s(T_g + 1)}{1 - T_s^2 T_g^2}$$

The absorption in each layer is then given by

$$Abs_s^{1sun} \approx I_{sens}(1-T_s)\frac{1+T_s T_g^2}{1-T_s^2 T_g^2} \quad Abs_g^{1sun} \approx I_{sens}(1-T_g)\frac{T_s(T_g+1)}{1-T_s^2 T_g^2}$$

The pump efficiency, which is the ratio between the gain media absorption and self-absorption, can be defined as

$$Abs_g^{1sun}/Abs_s^{1sun} \approx \frac{(1-T_g)}{(1-T_s)}\frac{T_s(T_g+1)}{1+T_s T_g^2} \sim \frac{(1-T_g)}{(1-T_s)} \sim \frac{\alpha_g t_g}{\alpha_s t_s}$$

As expected, the sum of the absorbed photon flux equals the excitation flux:

$$Abs_g^{1sun}(\theta, \lambda) + Abs_s^{1sun}(\theta, \lambda) \approx I_{sens}(\theta, \lambda)$$

To obtain the total absorption in each layer, $Abs_x^{1sun}(\theta, \lambda)$ are integrated over the emission of the sensitizer, polarization and angle $0 \leq \theta_s \leq \frac{\pi}{2}$.

**III-Modeling the emission of the sensitizer, including the directional emission and polarization**

Because we discriminate between the forward and backward direction of emission, the full solid angle is $2\pi$. Utilizing the Jacobian in spherical coordinates with full azimuthal symmetry, we can

write the normalization condition for the sensitizer

$$\frac{1}{2\pi} \sum_{s,p\ polarization} \sum_{emission} \int_0^{\pi/2} \int_0^{2\pi} I_{sens}(\lambda,\theta_s) \sin\theta_s\, d\varphi d\theta_s d\lambda$$

$$= \sum_{s,p\ polarization} \sum_{emission} \int_0^{\pi/2} I_{sens}(\lambda,\theta_s) \sin\theta_s\, d\theta_s d\lambda = \eta_s Abs_{sens}$$

Here $\varphi$ is the azimuthal angle in the sensitizer and $Abs_{sens}$ is the absorbed pump flux in the sensitizer. Therefore, the radiation emitted into the angle $d\theta_s$ for each polarization is $\frac{1}{2} I_{sens}(\lambda,\theta_s)\sin\theta_s$, where $\frac{1}{2}$ stands for the uniform distribution of polarizations and $0 \leq \theta_s \leq \frac{\pi}{2}$. Choosing isotropic emission, we can write that $I_{sens}(\lambda,\theta_s) = \eta_s Abs_s^{1sun} f(\lambda)$, where $f(\lambda)$ is the normalized emission of the sensitizer, i.e., $\int f(\lambda)d\lambda = 1$. In the general case, as long as the normalization conditions hold for $I_{sens}(\lambda,\theta_s)$, we can choose it to describe the emission with various angles and spectral distributions.

For instance, in order to describe oriented dipoles emission, titled with angle $\theta_d$ relative the z-axis, one may assume $I_{sens}(\lambda,\theta_s) = I_0 \sin^2(\theta_s - \theta_d) f(\lambda)$, where $I_0$ is a constant defined by normalization condition:

$$\sum_{s,p\ polarization} \sum_{emission} \int_0^{\pi/2} I_0 \sin^2(\theta_s - \theta_d) f(\lambda) \sin\theta_s\, d\theta_s d\lambda = \eta_s Abs_{sens}$$

Integrating over $\theta_s, \lambda$ we obtain $\sum_{s,p\ polarization} I_0 \left(\frac{1}{3} + \frac{1}{3}\cos^3\theta_d - \frac{1}{3}\sin 2\theta_d\right) = \eta_s Abs_{sens}$ which is used to define $I_0$. This approach also allows to describe polarized emission with coupling between the spectral properties and the spatial distribution in which $I_{sens}(\lambda,\theta_s)$ is not separable function of $\lambda, \theta_s$, as in the case of combination of different dyes with different orientations.

**IV-Modeling the complex angle of incidence**

In absorbing medium, the refractive index has a non-zero imaginary part. Therefore, the angle of incidence is complex and no longer represents the direction of propagation. However, most optical materials are weakly absorbing, i.e., $\left|Im(n_s)/Re(n_s)\right| = \left|\kappa_s/Re(n_s)\right| \ll 1$, where $\kappa_s = -Im(n_s) = \lambda \frac{\alpha_s}{4\pi}$ is the extinction coefficient of the sensitizer. In this case, the real part of the angle of incidence can be used as an approximation for the direction of the Poynting vector and, hence, the direction of the wave propagation. This condition holds true as long as $\lambda\alpha_s < 1$, or $\alpha_s < 10^6 m^{-1}$ at the sensitizer emission wavelength. We imply that the transverse component of the wave-vector is real[2-6], i.e., $Im(n_s k_0 \sin\theta_s^c)$, where $k_0$ is the free space wavenumber and $\theta_s^c$ is the proper complex-valued angle of incidence in the sensitizer. This arrangement requires a correction for our choice of real $\theta_s$ by setting $\sin\theta_s^c = \sin\theta_s + i\gamma_s$, where $\gamma_s = \frac{\kappa_s}{Re(n_s)} \sin\theta_s$ is the imaginary part of $sin\theta_s^c$. By Snell's law, we find that $n_g sin\theta_g^c = n_s sin\theta_s^c$, where $\theta_g^c$ is the proper complex-valued angle of incidence in the gain medium and $n_g$ is the complex refractive index of the gain medium, from which the real angle of incidence in the gain media is deduced: $sin\theta_g = Re(sin\theta_g^c) = Re\left(\frac{n_s}{n_g} sin\theta_s^c\right)$. In this case, we have $cos\theta_s = \sqrt{1 - sin^2\theta_s}$, $cos\theta_g = \sqrt{1 - sin^2\theta_g}$. Utilizing this approach, the single path absorption is calculated by the Beer-Lambert Law as $T_g(\lambda, \theta_g) = exp(-\alpha_g t_g / cos\theta_g) = exp(-\alpha_g t_g / \sqrt{1 - sin^2\theta_g})$ and $T_s(\lambda, \theta_s) = exp(-\alpha_s t_s / cos\theta_s) = exp(-\alpha_s t_s / \sqrt{1 - sin^2\theta_s})$.

**V-Second-order absorption and re-emission effects**

We treat the absorption and re-emission events in a similar way as for luminescent solar concentrators[7]. Noting that the rate of absorbed photons in each layer is proportional to the overall sensitizer emission $Abs_g^{1sun}, Abs_s^{1sun} \propto I_{sens}$, we can define the fraction of absorbed sensitizer emission in each layer by $Pabs_g = {Abs_g^{1sun}}/{I_{sens}}$, $Pabs_s = {Abs_s^{1sun}}/{I_{sens}}$. These values define the probability of the photon emitted by the sensitizer to be absorbed either in the sensitizer $Pabs_s$ or in the gain media $Pabs_g$. Each photon that is absorbed in the sensitizer is re-emitted with a probability of $\eta_s$ and can then be absorbed in the gain media with a probability of $Pabs_g$. Therefore, the rate at which photons are absorbed in the gain media is a result of the sum of the infinite series of first (I), second (II) and higher order emission events, i.e., $Pabs_g^{tot} = Pabs_g + (Pabs_s \eta_s)^1 Pabs_g + (Pabs_s \eta_s)^2 Pabs_g + \cdots = Pabs_g \sum_{i=0}^{\infty}(Pabs_s \eta_s)^i = \frac{Pabs_g}{1 - Pabs_s \eta_s}$, where $Pabs_g^{tot}$ is the probability of the photon that is emitted by the sensitizer to be absorbed in the gain medium after all of the absorption and re-emission events. These secondary effects become significant when the self-absorption and quantum efficiency are both high.

**VI – Guided modes loss**

Additionally, in the thin planar waveguide, the lasing mode tail overlaps with the high absorption region of the sensitizer – Figure **S2a** presents the fundamental (blue line) and first mode (red line) intensity in a planar waveguide with a core thickness of $t_{core} = 2\ \mu m$, wavelength $\lambda = 1064\ nm$, indices $n_{core} = 1.82$ and $n_{clad} = 1.7$ of the core and cladding, respectively. Figure **S2b** shows how the mode intensity propagating in the cladding region decreases with the core thickness (blue dashed and red dashed lines for the fundamental and first mode, respectively, left logarithmic axis), and how it affects the resonator Q-factor if the losses in the cladding are orders of magnitude higher than in the core. The perturbation theory for a low-loss approximation[8] gives a simple power average expression for a total attenuation coefficient $\alpha = \sum \alpha_i P_i / \sum P_i$. Here, $P_i$ is the fraction of the total power in each region obtained from the modal solution, and $\alpha_i$ is the corresponding attenuation coefficient, from which a $Q$-factor is deduced as $Q = \tau_c \nu_L$, where $\nu_L$ is the lasing frequency and $\tau_c$ is the photon lifetime in the cavity. The coherence time $\tau_c$[9-11] is given by $\tau_c^{-1} = \frac{c}{n}\left(\alpha - \frac{1}{l}ln\sqrt{R_{oc}}\right)$, with $n$ being an effective refractive index of the guided mode and $R_{oc}$ is the output coupling mirror's reflectivity. As an example, Figure **S2b** shows the resulting $Q$-factor for $\alpha_{core} = 0.3 m^{-1}$, $\alpha_{clad} = 10^3 m^{-1}$ (blue solid and red solid line for the fundamental and first mode, respectively, right logarithmic axis). As observed, even if a small fraction of the power propagates in the sensitizer, the $Q$-factor is limited; therefore, it is constructive to induce a spatial separation between the sensitizer and the gain medium layers. In contrast to near-field sensitization, radiative energy transfer allows avoiding this negative effect by introducing lossless cladding with an intermediate refractive index between the sensitizer and gain medium (as shown in Figure **1a**), which effectively confines the lasing mode in the low loss region and increases the Q-factor to the value of the unperturbed cavity. For this case, the results obtained in Figure **2** in the manuscript are valid as all simulations assumed lossless cladding with intermediate refractive index $n_{clad} = 1.75$.

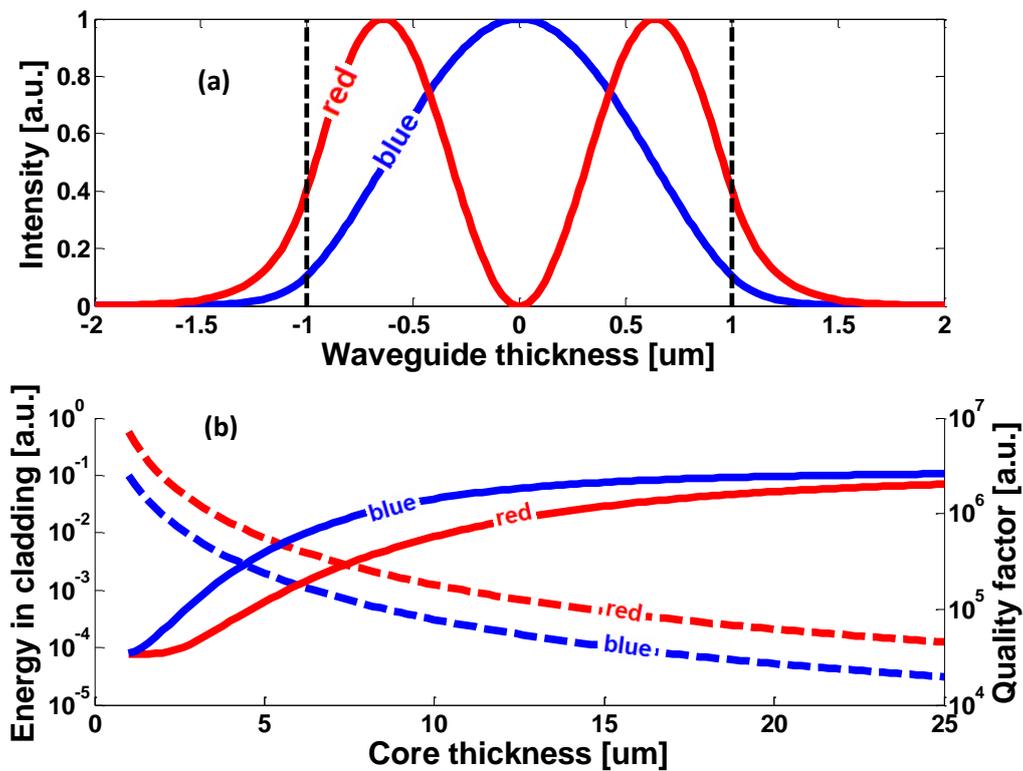

**Figure S2** (**a**) Intensity profile of guided modes in a symmetric slab waveguide with indices $n_{core} = 1.82$, $n_{clad} = 1.7$ of the core and cladding, respectively, for the core thickness $t_{core} = 2 \mu m$ (marked as vertical dashed black lines) and wavelength $\lambda_L = 1064\ nm$. The fundamental and first modes are shown in blue and red lines, respectively. (**b**) Fraction of the mode energy that propagates in the cladding as a function of the core thickness for the fundamental mode and first mode (blue and red dashed lines, respectively, left logarithmic axis), and the resulting attenuation of the Q-factor for the fundamental and first mode (blue and red solid lines, respectively, left logarithmic axis) for $\alpha_{core} = 0.3\ m^{-1}$, $\alpha_{clad} = 10^3\ m^{-1}$.

## VII-Sensitizer Intensity Response

To verify that AlQ$_3$:DCJTB(2%):Pt(TPBP)(4%) makes it possible to construct a SPL we tested glass slides with deposited AlQ3:DCJTB(2%):Pt(TPBP)(4%) under various solar concentrations. Red line in **Figure S3** shows that the luminescent output of the sample starts to decrease already at concentrations above 1 sun. The values in blue line of **Figure S3** are normalized to the linear extrapolation of the response obtained under very low solar concentrations.

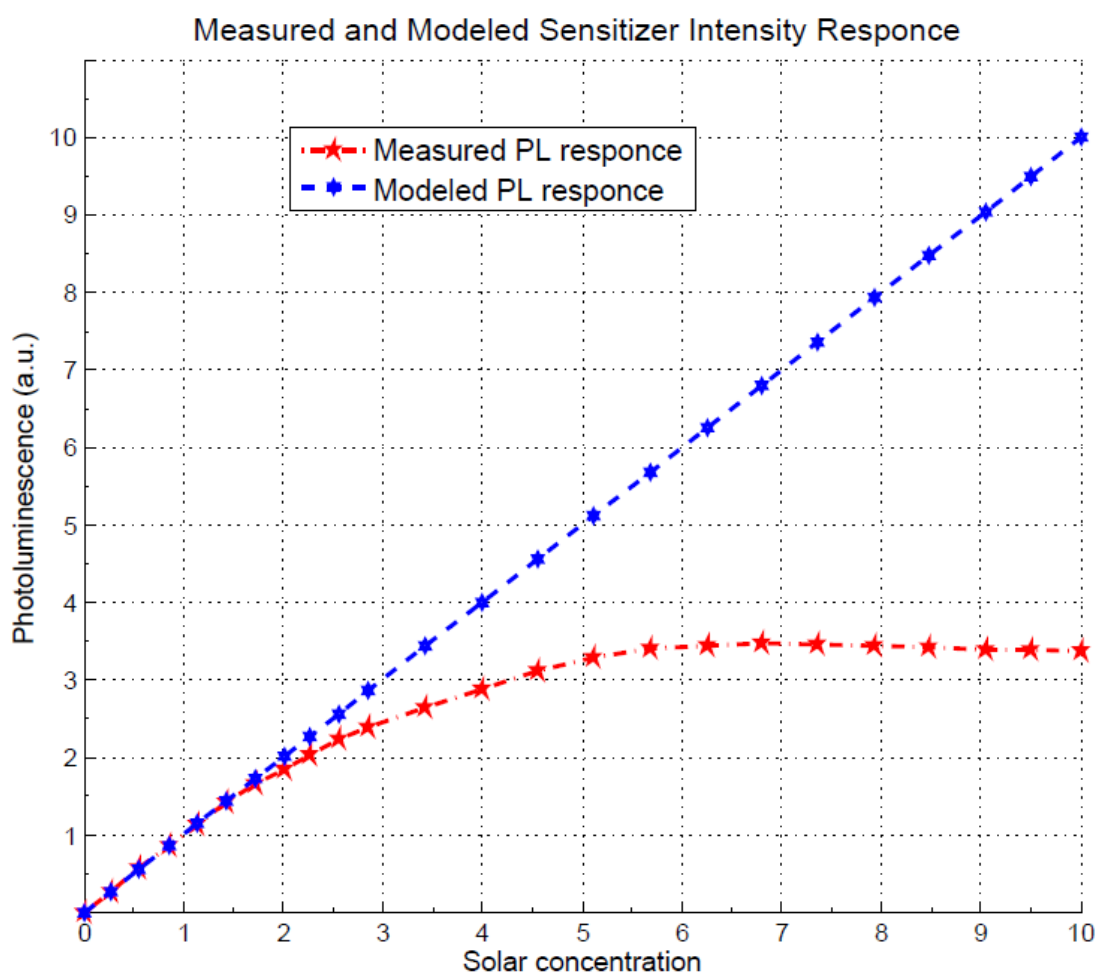

**Figure 3** Linear extrapolation of the intensity response of AlQ$_3$:DCJTB(2%):Pt(TPBP)(4%) obtained at very low solar concentrations (blue) and actual behavior of this organic complex when illuminated with solar radiation at different concentrations.

# References

1. Siegel, R. Net radiation method for enclosure systems involving partially transparent walls. (1973).

2. Orfanidis, S. J. Electromagnetic Waves and Antennas. (§7.9). Available at: http://www.ece.rutgers.edu/~orfanidi/ewa/.

3. Parmigiani, F. Some aspects of the reflection and refraction of an electromagnetic wave at an absorbing surface. Am. J. Phys. **51,** 245–247 (1983).

4. Roo, R. D. & Tai, C.-T. Plane wave reflection and refraction involving a finitely conducting medium. IEEE Antennas Propag. Mag. **45,** 54–61 (2003).

5. Chang, P. C. Y., Walker, J. G. & Hopcraft, K. I. Ray tracing in absorbing media. J. Quant. Spectrosc. Radiat. Transf. **96,** 327–341 (2005).

6. Dupertuis, M. A., Acklin, B. & Proctor, M. Generalization of complex Snell–Descartes and Fresnel laws. J. Opt. Soc. Am. A **11,** 1159 (1994).

7. Batchelder, J. S., Zewai, A. H. & Cole, T. Luminescent solar concentrators. 1: Theory of operation and techniques for performance evaluation. Appl. Opt. **18,** 3090–3110 (1979).

8. Adams, M. J. Loss calculations in weakly-guiding optical dielectric waveguides. Opt. Commun. **23,** 105–108 (1977).

9. Grivas, C. Optically pumped planar waveguide lasers, Part I: Fundamentals and fabrication techniques. Prog. Quantum Electron. **35,** 159–239 (2011).

10. Grivas, C. Optically pumped planar waveguide lasers: Part II: Gain media, laser systems, and applications. Prog. Quantum Electron. **45–46,** 3–160 (2016).

11. Yariv, A. Photonics : optical electronics in modern communications /. (Oxford University Press, 2007).